\documentclass[runningheads]{llncs}
\input{preamble.sty}

\newif\ifanonymous
\anonymousfalse

\ifanonymous
\author{Anonymous Author(s)}
\institute{No Institute Given}
\else
\author{Wander Nauta\orcidlink{0009-0009-7521-6408} \and
        Marcus Gerhold\orcidlink{0000-0002-2655-9617} \and
        Marieke Huisman\orcidlink{0000-0003-4467-072X}}
\authorrunning{W. Nauta et al.}
\institute{University of Twente, Formal Methods and Tools, Netherlands}
\fi

\title{Crash-free Deductive Verifiers\thanks{This is a preprint. The Version of Record is published in \emph{Model Checking Software (SPIN 2026)} and is available at TBD.}}
\begin{document}

\maketitle   

\begin{abstract}
As deductive verifiers mature, their potential user base is growing from the initial core developers to other users.
To convince external users of the suitability of verifiers, these tools must run reliably out of the box, give meaningful error messages and display correct results.
Yet deductive verifiers are large and complex software systems and their own full verification is often out of reach.
We therefore need complementary means to provide such guarantees.
This paper advocates the use of fuzzing as a practical way to improve the quality and robustness of deductive verifiers.
We outline how fuzz testing can be applied to deductive verifiers, and demonstrate the idea with the prototype tool \avalanche, which is integrated with the \V/ verifier.
We report on our experiments in which \avalanche uncovered several issues in \V/ and demonstrate that the approach also works for other deductive verifiers.

\end{abstract}
\section{Introduction}
\label{sec:intro}

Over the last years, we have seen steady progress in automated formal analysis tools that help software developers to increase  confidence in the quality of their software. To enable software developers to use these tools efficiently, the tools themselves must be robust: they should not crash, they should not slow down development, and they should give correct outputs. Many formal analysis tools are developed in an  academic setting, where ensuring their robustness is not always a first priority.
We argue that it is important to provide an approach---supported by suitable tools---that enables authors of formal analysis tools to quickly identify bugs, for example in their front-ends. Having this support would free up time for developers to focus on the formal analysis aspects themselves.

This paper advocates the use of fuzzing as practical means to make deductive verifiers more robust and reliable. Fuzzing is a form of automated testing that generates random inputs for a program under test. We show how to use fuzzing to easily identify inputs, consisting of both programs and their contracts, that make a deductive verification tool crash. In particular, we demonstrate the approach with our prototype tool \avalanche\footnote{\ifanonymous Tool documentation will be made available after the anonymous reviewing process.\else This work is based on the thesis of Wander Nauta~\cite{Nauta2025}. \A/'s source code and documentation are available at \url{https://github.com/utwente-fmt/vercors}.\fi}, which integrates fuzz testing into the VerCors verifier~\cite{armborstvercorsverifierprogress2024}. With \avalanche, we specifically target the robustness of \V/ \emph{itself}, i.e. verifier crashes of VerCors, using different fuzzing strategies. These are errors that can be traced to bugs within \V/, i.e., errors occurring after the (ANTLR-generated~\cite{parrANTLRPredicatedLL1995}) parser, but before the (Viper-provided~\cite{mullerViperVerificationInfrastructure2016}) back-end. We emphasize that we do not target unsoundness or incompleteness issues of VerCors, and we ignore (well-defined) error messages about undefined features. 
Finally, we show that the fuzzing approach itself is tool-independent, and that it can also uncover front-end bugs in other deductive verifiers, such as Dafny~\cite{leinoDafnyAutomaticProgram2010}, VeriFast~\cite{VeriFast}, and Viper~\cite{mullerViperVerificationInfrastructure2016}.  


The remainder of this paper is organised as follows: Section~\ref{sec:background} briefly introduces fuzzing and the \vercors verifier.
Section~\ref{sec:tool} presents our approach and its implementation in the prototype tool \avalanche. It also demonstrates the usability of \avalanche by applying it to VerCors and discusses how the approach has been used on several other verifiers. 
Section~\ref{sec:related} discusses related work and Section~\ref{sec:concl} concludes the paper with outlining future work.


\section{Background and Motivation}
\label{sec:background}
\paragraph{The VerCors Verifier.}
The \V/ verifier is a \emph{deductive program verifier}~\cite{armborstvercorsverifierprogress2024} that is based on permission-based separation logic \cite{haackPermissionBasedSeparationLogic2015}, which is an adaptation of concurrent separation logic \cite{OHEARN2007271}. VerCors proves memory safety, data race freedom and functional correctness for a number of programming languages, including Java, \CPP/, C, and OpenCL~\cite{armborstvercorsverifierprogress2024}.
It is targeting concurrent programs in particular that are either using explicit threads or GPU-based parallelism.


\V/ works statically, without executing the program that is being verified. Annotated programs are encoded into the Viper framework~\cite{mullerViperVerificationInfrastructure2016}, which then uses symbolic execution or weakest precondition generation to compute a number of proof obligations, which are then discharged by Z3~\cite{demouraZ3EfficientSMT}. 

\paragraph{Fuzzing and Test Case Generation.}
\emph{Fuzzing} or \emph{fuzz testing} is an automated testing approach where many test inputs are generated (semi-)randomly and executed in the hope of uncovering incorrect behaviour such as crashes \emph{by chance}~\cite{purdomSentenceGeneratorTesting1972,millerEmpiricalStudyReliability1990}.
In its purest form, fuzzing rarely executes a program's deep functionality, as most random inputs are invalid, and will be immediately rejected.
Also, due to the infinite input space it is impossible to predict when and if the fuzzer will find a next crash.
Therefore, modern fuzzers 
provide different fuzzing strategies, such as coverage guidance or grammar-based methods to execute deeper functionality of programs, cf. Man\`es et al.~\cite{manesArtScienceEngineering2021} for an extensive survey. 


As our goal is to develop testing techniques for program verifiers that take (specified) programs as input, fuzzing techniques for compilers are of particular interest for us. 
\emph{Grammar-based fuzzing} ensures that generated inputs conform to a language, and therefore pass parsing.
The fuzzer \emph{Grammarinator} by Hodov\'an et al. \cite{hodovanGrammarinatorGrammarbasedOpen2018}, for example, generates syntactically correct inputs from ANTLR grammars, which helps to expose bugs in the resolution or rewriting phases.


To trigger even `deeper' bugs, we use randomized test-case generation tools as  fuzzers, which produce \emph{semantically} valid programs.
For instance, Csmith~\cite{yangFindingUnderstandingBugs2011} has been successful in finding bugs in C compilers.
It acts like a grammar-based fuzzer, but 
generates only a subset of programs that are semantically valid and have known, safe and defined behaviour.
Going beyond C and to \V/' Prototypal Verification Language (PVL) our work integrates Xsmith~\cite{hatchGeneratingConformingPrograms2023} as a framework that supports language-specific fuzzer generation. 


%
%

\section{Fuzzing for Deductive Verifiers}\label{sec:tool}
To easily identify (and fix) inputs that make a deductive verifier crash, we advocate the use of fuzzing. We discuss different fuzzing strategies that can be used for this. We  outline how we implemented \avalanche to set up this approach for the \vercors verifier, and we discuss our experimental setup and performance metrics to evaluate \avalanche on \vercors. Finally, we discuss some experiments that demonstrate that the approach also identifies errors in other deductive verifiers.
\vspace{-1em}
\subsection{Fuzzing Strategies}
\label{sec:strats}
\vspace{-0.5em}
As deductive verifiers take (annotated) programs as input, in the literature various fuzzing strategies have been proposed that are suitable for this use case: 
\medskip\\
\noindent \textbf{Coverage-guided fuzzing} (without grammars) is given by the Jazzer fuzzer~\cite{codeintelligenceJazzerCoverageguidedInprocess}. Jazzer is an
    in-process fuzzer that instruments classes on-the-fly using logic from the JaCoCo\footnote{\url{https://www.eclemma.org/jacoco/}} code coverage
    library and mutation logic from LLVM's libFuzzer\footnote{\url{https://llvm.org/docs/LibFuzzer.html}}. 
    This language-agnostic approach does not rely on verifier-specific knowledge, and therefore serves as a baseline for comparison with the other strategies. 
    %
    
\noindent \textbf{Grammar-based (syntactic)} generation is supported via
    Grammarinator~\cite{hodovanGrammarinatorGrammarbasedOpen2018}. Its goal is to generate inputs that randomly cover the entire grammar of the input language.
    Grammarinator takes an ANTLR~\cite{parrANTLRPredicatedLL1995} grammar
    as input and generates a fuzzer based on it. 
    All generated inputs are grammatically correct by construction and therefore pass parsing. 
    However, the grammar-based fuzzer is not aware of the semantics of the
    programs it generates, the scopes of variables, or the types of
    expressions, which can limit the usability of this strategy.

\noindent\textbf{Grammar-based (syntactic \& coverage)} generation combines Grammarinator grammars with coverage feedback, using Jazzer-generated input as seed. 
 
\noindent\textbf{Grammar-based (verifiable subset)} generation uses  Xsmith~\cite{hatchGeneratingConformingPrograms2023}. 
Xsmith generates input that can be parsed and type-checked, and that could successfully be passed to a backend verifier if enabled.
 

\subsection{\avalanche Implementation}
To demonstrate the importance of setting up a good fuzz testing approach for deductive verifiers, we developed the tool \avalanche. 
It is designed to be agnostic to the fuzzing strategy that is employed: it currently can work with Jazzer, Grammarinator and Xsmith to generate input, but it is easy to replace these with any future strategy implementation.
To tailor \avalanche to VerCors, we ported the PVL, C, \CPP/, and Java ANTLR grammars from \V/ to Grammarinator. Since Xsmith grammars are more involved and ANTLR grammars cannot be directly converted into them, \avalanche currently only generates PVL inputs using Xsmith.
As an example of \A/'s extensibility, it also integrates the general-purpose blind fuzzer Radamsa\footnote{\url{https://gitlab.com/akihe/radamsa}}, primarily to validate the generator interface. We will not use it later in the comparative evaluation.

\A/ continuously runs \V/ on inputs from a fuzzer, and watches for crashes, which---if found---are collated to remove duplicates. 
Whenever a new crashing input is discovered, it is documented with relevant information, such as the stack trace and version details, to enable efficient triaging.
\avalanche also includes a web interface for monitoring ongoing fuzzing campaigns. 

If a crashing input is found by \avalanche, it is likely that an
infinite number of potential inputs would trigger what is essentially
the same crash. 
\A/ attempts to combine crashing inputs by inspecting the stack trace of the exception that caused the crash. 
For each stack frame, a hash is computed over the 
class name, method name, file name, and line number.
Crashes that hash to the same value are treated as duplicates.
This entire workflow is automated: \avalanche  repeatedly runs \V/ on generated input programs, detects and classifies crashes, and reports new distinct failures to developers.
\subsection{Experimental Results for \avalanche}
\label{sec:eval}
%
%

%
We compare the different fuzzing strategies that are supported in \avalanche when it is applied to \V/, using code coverage over time as measure. 
Concretely, we measure the number of instrumentation points, i.e., code locations dynamically inserted by Jazzer through the JaCoCo library to track execution coverage, 
in a fixed amount of time, on the same environment, and on the same machine. This provides a practical proxy for how much of the verifier’s code is exercised by each fuzzing strategy, and is standard in evaluating coverage-guided fuzzers~\cite{manesArtScienceEngineering2021,kleesEvaluatingFuzzTesting2018}.
We found that a time limit of 5 minutes was already sufficient to see a marked performance difference between the different approaches.



All measurements are done on a Lenovo ThinkPad P16v Gen 2, Intel Core Ultra 7 155H (x86\_64), running NixOS, with \V/ development version \texttt{2bd3bcaed} (based on released version 2.2.0).
The Linux CPU performance governor is set to `performance', clocked at a maximum of 4.8 GHz.
We avoid exercising code that is not part of \V/ by using the \texttt{-{}-skip-backend} flag, so the Viper back-end is never invoked.
Time is measured from the start of the process and thus includes a short startup delay. 

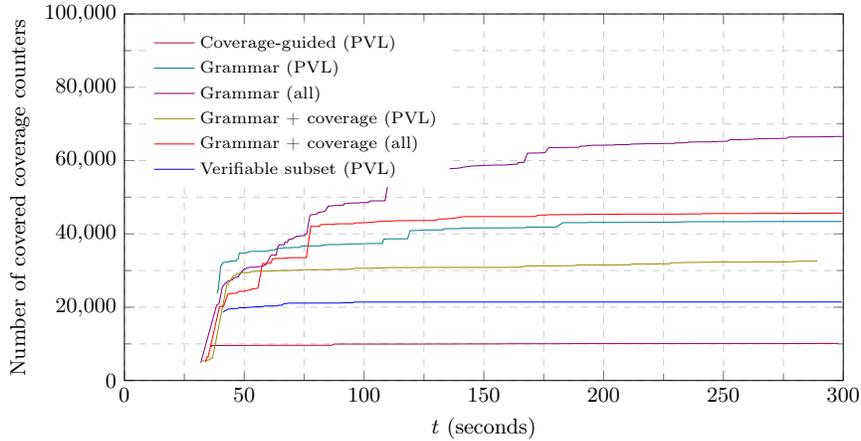
\begin{figure}
    \begin{tikzpicture}
    \scalebox{0.9}{
        \begin{axis}[
        xmax=300,
        ymax=100000,
        width=\linewidth,
        height=7cm,
        xlabel={$t$ (seconds)},
        legend pos=north west,
        ylabel={Number of covered coverage counters},  
        legend style={font=\scriptsize},
        grid=both,
        minor tick num=1,
        grid style={dashed, gray!40}
    ]
 
            \addlegendentry{Coverage-guided (PVL)}
            \addlegendentry{Grammar (PVL)}
            \addlegendentry{Grammar (all)}
            \addlegendentry{Grammar + coverage (PVL)}
            \addlegendentry{Grammar + coverage (all)}
            \addlegendentry{Verifiable subset (PVL)}
               \addplot+ [solid,purple,mark=none] table {dat/direct_pvl_1_covered.dat};
    \addplot+ [solid,teal,mark=none] table {dat/gram_pvl_1_covered.dat};
    \addplot+ [solid,violet,mark=none] table {dat/gram_all_1_covered.dat};
    \addplot+ [solid,olive,mark=none] table {dat/gui_pvl_1_covered.dat};
    \addplot+ [solid,red,mark=none] table {dat/gui_all_1_covered.dat};
    \addplot+ [solid,blue,mark=none] table {dat/smith_pvl_1_covered.dat};
\end{axis}
}
\end{tikzpicture}
    \caption{Performance of the different approaches compared.}
    \label{fig:all}
\end{figure}
Figure~\ref{fig:all} presents the combined measurements and shows the growth of covered coverage counters over time. We distinguish grammars limited to PVL and those combining all front-end languages (PVL, Java, C, and C++) where the latter explore a broader syntactic space but follow similar qualitative trends.

%
The direct approach (coverage-guided fuzzing) did not find any bugs. It achieves low and barely growing coverage, indicating that this approach is not able to find interesting inputs. 
The number of instrumented coverage counters also does not grow, thus a large part of the \V/ codebase is not exercised by the generated inputs. Manual inspection of the generated inputs shows that the fuzzer primarily finds inputs that are
immediately rejected by the PVL parser. 

Grammar-based fuzzing generates  inputs that are varied enough to find and instrument new parts of the \V/ codebase. 
The number of paths covered also grows, albeit more slowly. 
Purely syntactic grammar-based fuzzing finds errors in the 
resolution or rewriting phase, but when adding coverage, 
we also find errors in the resolving stage of \vercors.


Adding coverage feedback (i.e. using Jazzer) does not change the kinds of errors found and slightly reduces performance: the Grammar (PVL) curve (green) reaches higher coverage than the corresponding Grammar + coverage (PVL) curve (yellow) which indicates that overhead outweighs improved exploration.

The Xsmith-based generation of a verifiable subset\footnote{which only works for PVL currently.} starts off strong: as we generate only programs that can be parsed and resolved, the initial test (at $t=40$) already exercises a significant part of the PVL front-end and a number of rewrite phases, which together cover a sizable chunk of the codebase.
However, the coverage  does not increase much beyond the initial attempts.

The errors that we found in our experiments in general are relatively shallow: the crashing inputs are relatively short after minimization, and the cause of the crash is clear from the input and stack trace. 
Appendix~\ref{sec:allbugs} gives an overview of all issues that were reported to the \vercors developers.
For example, errors that we found were \vercors crashing on: an empty `enum' block (issue 1248), 
annotations that were written in a wrong location, which was accidentally admitted by the grammar (issues 1302 and 1261), 
legal program expressions that were not covered by \vercors, but also not properly handled
(issue 1302). 
Several issues were found that would have been unlikely to be caught in either code review or tests, such as 
a parse error that only appeared in the \CPP/ front-end of \vercors, and only for names that consisted \emph{entirely} of the letters u and l
(issue 1294). 

\paragraph{Measurements on other Verifiers.}
\label{sec:other}
%
We have also investigated 
whether the coverage-guided and grammar-based fuzzing approaches can discover bugs in verification tools. 
Using a coverage-guided, not grammar-based, fuzzer, we found three issues in VeriFast~\cite{VeriFast}, two of which were new and had not yet been reported. For the Carbon and Silicon verifiers---both part of the Viper project from the ETH Z\"urich \cite{mullerViperVerificationInfrastructure2016}---we used both coverage-guided and grammar-based fuzzing and uncovered an already reported bug. 
Finally, we applied our approach to Dafny~\cite{leinoDafnyAutomaticProgram2010}, but did not uncover any issues. One reason may be that Dafny already benefits from several dedicated testing techniques developed in earlier work, see e.g.~\cite{irfanTestingDafnyExperience2022}. 

\section{Related Work}\label{sec:related}
Man\`es et al.~\cite{manesArtScienceEngineering2021} provide an extensive survey and taxonomy of fuzzers. Here, we focus on prior efforts that apply fuzzing  to program verification tools. 
Irfan et al. \cite{irfanTestingDafnyExperience2022} built XDsmith, which uses Xsmith~\cite{hatchGeneratingConformingPrograms2023} to generate syntactically correct, annotated Dafny programs 
that have a verification outcome that is already known during generation. 
This allows XDsmith to test whether the verifier produces the expected verification results.
Thus, their focus is on the \emph{correctness} of the verification engine, rather than on the tool's \emph{robustness} with respect to 
inputs.

Two other fuzzers for Dafny exist:
fuzz-d \cite{usherFuzzdRandomProgram2023} and DafnyFuzz
\cite{donaldsonRandomisedTestingCompiler2024}. 
Their primary focus is to look for miscompilation, where the translation to the target language changes the program's semantics. 
Both tools employ differential testing, comparing multiple compilation targets, and can also use self-checking oracles to validate expected outcomes.
In comparison to our work, the exposure of Dafny's compiler front-end is a by-product, and not the main focus.

\section{Conclusions}
\label{sec:concl}
%
We demonstrated that fuzzing is an effective and practical approach to test the \emph{robustness} (rather than the \emph{correctness}) of deductive verifiers. 
Our prototype tool, \A/, applies different fuzzing strategies to the \V/ verifier and has revealed several issues. 
We showed the extensibility of the approach to other deductive verifiers and note that the variety of fuzzing strategies was essential to identify different issues. 

As future work, we aim to improve the usability of \A/ by supporting automatic minimization of crashing inputs. This makes test cases easier for developers to understand and reproduce.

\bibliographystyle{splncs04}
\bibliography{thebib}

\clearpage
\appendix
\section{Overview of errors found with \avalanche}
\label{sec:allbugs}
This appendix is included for the reviewing process. If the paper is accepted, we plan to make this list available online.

\section*{Issues discovered}
The following issues in \V/ were discovered using different versions of the \A/ tool.
The issues are listed in chronological order of discovery and are
numbered for easy reference with the \V/ repository \ifanonymous
\emph{link blinded for review}\else, \url{https://github.com/utwente-fmt/vercors}\fi.

In each case, a (manually) minimized input is given as an example, rather than the complete input as generated by the fuzzer. The inputs are not intended to be examples of actual, useful programs, but they are grammatically correct.

The repository's issue tracker has more details for each issue, including a change history pointing at the \V/ version where an issue was fixed, if applicable.

\subsection*{Issue 1248: Enum with no members crashes \V/}

The following PVL input program caused \V/ to crash:

\begin{lstlisting}
enum Empty {
}
\end{lstlisting}

\noindent This program is now rejected with a parse error.

\subsection*{Issue 1260: All-underscores name crashes \V/} \label{i1260}

Splitting and combining names consisting of only underscores, such as in the following PVL program, caused \V/ to crash:

\begin{lstlisting}
void ___() {
}
\end{lstlisting}

\noindent This has been resolved by passing along such names as-is.

\subsection*{Issue 1261: Unexpected bip\_annotation keyword crashes \V/} \label{i1261}

The \emph{bip\_annotation} keyword is specific to the JavaBIP verification feature \cite{bliudzeJavaBIPMeetsVerCors2023}, but it was also parsed in the C and PVL front-ends, where it caused a crash:

\begin{lstlisting}
bip_annotation
void foo() {
}
\end{lstlisting}

\noindent The occurrence of the \emph{bip\_annotation} keyword is now correctly marked as an error.

\subsection*{Issue 1263: Label declarations outside expected scopes crash \V/} \label{i1263}

The PVL language has some constructs that are similar to method bodies, including the \emph{run} block, the \emph{constructor} block, and the \emph{vesuv\_entry} block.
In these blocks, labels (for use with the \emph{goto} statement) were syntactically allowed, but placing them there caused \V/ to crash.

\begin{lstlisting}
class Three {
    run {
        label sixty;
    }
}
\end{lstlisting}

\noindent This has been resolved.

\subsection*{Issue 1264: Special expressions outside channel invariant crash \V/} \label{i1264}

As part of the VeyMont \cite{rubbensVeyMontChoreographyBasedGeneration2025} support for verifying choreographies, \V/ supports `channel invariant expressions', specifically including the \emph{\textbackslash{}msg} (message), \emph{\textbackslash{}sender} and \emph{\textbackslash{}receiver} keywords.
These expressions only have a meaning inside a \emph{channel\_invariant} statement, but were also grammatically accepted in other places, where they caused a crash.

\begin{lstlisting}
requires \msg;
void foo() {
}
\end{lstlisting}

\noindent This is now correctly reported as an error.

\subsection*{Issue 1265: Type of \texttt{\textbackslash{}type}, \texttt{\textbackslash{}typeof} expressions is inconsistent} \label{i1265}

There is some support in the PVL and Java front-ends for referring to the types of expressions, to support the Java \emph{instanceof} operator among other features.
However, the type of these expressions is not documented, and different rewriting phases disagreed whether it should be some form of \emph{type<>} or an integer, which meant that the following programs both failed with an error message suggesting the other would succeed:

\begin{lstlisting}
class A {
}

type<A> foo() {
   return \type(A);
}
\end{lstlisting}

\begin{lstlisting}
class A {
}

int foo() {
   return \type(A);
}
\end{lstlisting}

\noindent The \emph{\textbackslash{}type} and \emph{\textbackslash{}typeof} expressions are not believed to be widely used.

\subsection*{Issue 1266: Right plus operator overload syntax error crashes \V/} \label{i1266}

PVL supports operator overloading, specifically for the plus (addition) operator.
This is intended for cases like the string class, where the plus operator is overloaded to represent string concatenation.
The class author has the option to decide whether the operator they define should be left-associative by defining \emph{+} or right-associative by defining \emph{right+}.
However, the PVL grammar also accepted other keywords, which would then cause a crash:

\begin{lstlisting}
class Th {
    ensures \result;
    bool wrong+(Th a) {
        return true;
    }
}
\end{lstlisting}

\noindent This is now an error.

\subsection*{Issue 1267: Returning a value of type \texttt{resource} crashes \V/} \label{i1267}

The \emph{resource} type is a `Boolean-like type' that is intended to be used with \V/ separating conjunction expression, a concept from separation logic. However, in some cases using this type with regular Boolean expressions in a \emph{return} statement would cause a crash:

\begin{lstlisting}
resource bar() {
    return true;
}

void foo() {
    assert bar();
}
\end{lstlisting}

\noindent This has been worked around by marking returning \emph{resource}s as temporarily unsupported.
The input no longer crashes.

\subsection*{Issue 1268: Neglecting to specify type variable crashes \V/} \label{i1268}

Attempting to create (using the \emph{new} expression) an instance of a generic type without specifying type variables would cause \V/ to crash:

\begin{lstlisting}
class X<T> {
}

void main() {
    new X();
}
\end{lstlisting}

\noindent This is now reported as a type error that says that the \emph{new} expression only supports non-generic classes; not being able to instantiate generic classes is therefore a language limitation, and no longer a crash.

\subsection*{Issue 1273: Generic classes with final fields crash \V/} \label{i1273}

A PVL program like the following would cause a crash when a rewriting phase caused the abstract syntax tree to no longer typecheck:

\begin{lstlisting}
class C <T> {
    final int f;
}
\end{lstlisting}

\subsection*{Issue 1290: Prover types and prover functions outside PVL crash \V/} \label{i1290}

The \emph{prover\_type} and \emph{prover\_function} spec keywords allow adding definitions directly to the underlying SMT solver, for instance defining a function directly in SMTLIB syntax.
This was only supported in PVL, but included in other language grammars as well, which meant that attempting to verify the following as e.g. C code would hit a match error:

\begin{lstlisting}
/*@ prover_function int two() \smtlib `(+ 1 1)`; */
\end{lstlisting}

\subsection*{Issue 1291: C union declarations are not yet supported} \label{i1291}

There is no support yet for unions in the C implementation.
Normally, this would count as a language limitation and so not as an error; however, in this case the omission causes a ParseMatchError, which does cause a crash in our definition.

\begin{lstlisting}
union onion {
    float goat;
};
\end{lstlisting}

\subsection*{Issue 1292: Non-method members of Java @interfaces not yet supported} \label{i1292}

Java annotation interfaces are allowed to declare nested classes, interfaces, enums and constrants, but this is not yet supported by \V/, also triggering a \emph{ParseMatchError}.
The following example from the Java Language Specification\footnote{\url{https://docs.oracle.com/javase/specs/jls/se23/html/}} shows a possible use of this language feature, which appears to be quite rarely used otherwise:

\begin{lstlisting}
@interface Quality {
    enum Level { BAD, INDIFFERENT, GOOD }
    Level value();
}
\end{lstlisting}

\subsection*{Issue 1293: Non-inline thread-local predicates are not rewritten} \label{i1293}

The \V/ specification language (here demonstrated using PVL) allows declaring and defining predicates: functions returning `resources', which are similar to booleans but are used to represent a read or write permission.
These predicates can optionally be marked \emph{inline}, \emph{thread\_local}, or both.
The case where a predicate was marked \emph{thread\_local} but not \emph{inline} was not handled by \V/, which would cause an error just before the program was handed off to the Silver backend:

\begin{lstlisting}
thread_local resource fox();
\end{lstlisting}

\subsection*{Issue 1294: \CPP/ front-end fails to parse specific names} \label{i1294}

A bug in the lexer grammar meant that the following input did not parse as \CPP/:

\begin{lstlisting}
namespace u {
}
\end{lstlisting}

\subsection*{Issue 1298: lock/unlock statement with literal null crashes \V/}

The \emph{lock} statement takes a non-null expression.
This is checked, but the check does not correctly handle a null literal, triggering a UnreachableAfterTypeCheck error.

\begin{lstlisting}
void example() {
    lock null;
}
\end{lstlisting}

\subsection*{Issue 1299: Empty sequential block triggers crashes \V/}

The \V/ toolset is especially geared towards verifying the correctness of concurrent programs.
The \emph{par} (parallel) and \emph{sequential} blocks in PVL allow specifying programs that may involve multiple threads and proving their correctness.
However, \V/ expects these \emph{sequential} blocks to be nonempty, triggering a `tail of empty list' exception.

\begin{lstlisting}
void x() {
    sequential {
    }
}
\end{lstlisting}

\subsection*{Issue 1300: fork/join statement with literal null crashes \V/}

The PVL \emph{fork} statement expects an object-typed expression of a class that has a run method, but the grammar also allows a literal \emph{null} keyword.
This passes at least some type checks, but then causes a ClassCastException in the code that checks whether the expression is a runnable.

\begin{lstlisting}
void spork() {
    fork null;
}
\end{lstlisting}

\subsection*{Issue 1302: `old' expression in context or requires clause crashes \V/}

The \emph{\char`\\old(expression)} syntax is `typically used in postconditions (ensures) or loop invariants'.
However, in PVL it is also grammatically accepted in other places where an expression is expected, which then causes \V/ to generate an abstract syntax tree that is rejected by the Viper back-end.
The rejection is treated as a crash.

\begin{lstlisting}
requires \old(true);
void example() {}
\end{lstlisting}

\subsection*{Issue 1303: Exponentiation/power expression crashes \V/}

In the documentation for PVL, an exponentiation (power) operator is described, consisting of two caret characters.
Attempting to use this operator causes \V/ to crash.

\begin{lstlisting}
void zap() {
    int i = 2 ^^ 3;
}
\end{lstlisting}

\subsection*{Issue 1304: Simplification rule in program crashes \V/}

\V/ contains a number of axioms (rules) that can be used to simplify inputs.
For instance, if an expression $i - i$ appears in a program, with $i$ some number, it can be simplified to $0$.
These rules are specified in PVL syntax in the file \emph{simplify.pvl}, which is bundled with \V/.
The issue here is that including a simplification rule in the program being verified (so outside \emph{simplify.pvl}) was accepted by the grammar but not correctly handled by \V/, causing a \emph{ColToSilver\$NotSupported} exception.

\begin{lstlisting}
axiom add { 2 + 2 == 4 }
\end{lstlisting}

\subsection*{Issue 1306: Type error in context\_everywhere crashes \V/}

The \emph{context\_everywhere} statement, which is a shorthand to introduce a single expression as a precondition, a postcondition, and a loop invariant for all loops in a method.
The expression is expected to be of type \emph{resource} (often a boolean).
If some other type of expression is given, this causes an internal type error, specifically an \emph{CoercingRewriter\$Incoercible} exception.

\begin{lstlisting}
context_everywhere 1;
void one() {
}
\end{lstlisting}

\subsection*{Issue 1307: Referring to sibling par inside parallel block can crash \V/}

The following PVL program, which uses the \emph{parallel}, \emph{par} and \emph{barrier} constructs incorrectly, caused \V/ to throw a \emph{TimeTravel} exception.

\begin{lstlisting}
void x() {
    parallel {
        par ty {}
        par barrier(ty) {}
    }
}
\end{lstlisting}

\subsection*{Issue 1308: Function/predicate arguments of type type<...> crash \V/}

The \emph{type<...>} generic type represents a type; it is the result of the \emph{\textbackslash{}typeof} operator (see also issue 1265).
However, declaring a function, method or predicate that accepts such a value triggers \emph{ColToSilver\$NotSupported}.
In PVL, this looks as follows:

\begin{lstlisting}
void x(type<void> y) {
}
\end{lstlisting}

\subsection*{Issue 1312: Postfix inc/dec without permission crashes \V/}

The following program should result in a permission error, since the `acters' method accesses the field `erior' without read or write permissions.
Instead, a \emph{BlameUnreachable} exception is thrown: some code asserts that `assigning to a field should trigger an error on the assignment, and not on the dereference', an assertion that apparently trips on this case as well.
The issue occurs in PVL and Java, but not in C or \CPP/.
In PVL, it looks as follows:

\begin{lstlisting}
class room {
    int erior;

    char acters() {
        erior++;
    }
}
\end{lstlisting}

\subsection*{Issue 1313: Label in Java initializer block crashes \V/}

In Java, classes may contain blocks of statements to initialize instance or static members.
These blocks are supported by \V/, but attempting to insert a label into such a block would trigger a \emph{Scopes\$NoScope} exception.
This issue is similar to issue 1263, above, but in the more rarely used initializer block context.
An example input is:

\begin{lstlisting}
class Z {
    { m:; }
}
\end{lstlisting}

\subsection*{Issue 1316: string keyword in C and C++ spec crashes \V/}

The C and \CPP/ string types (\emph{char*} and \emph{std::string}, respectively) are not yet well-supported by \V/.
However, for both languages, the \emph{string} keyword is recognized in the grammar, apparently referring to some specification-level string type that has not been implemented yet.
When an input containing a specification-level \emph{string} keyword is parsed and converted to COL, this results in a \emph{ParseMatchError}, as follows:

\begin{lstlisting}
//@ pure string cheese() = "mozzarella";
\end{lstlisting}

\subsection*{Issue 1330: Names with many trailing digits crash \V/}

In some instances, \V/ will append or strip a trailing numeric suffix from a declaration or label name to disambiguate between different instances.
Some code in the \emph{Namer} class assumed that such a suffix would always be parseable as a Java Integer, but this is not the case.
A suffix that is too large would raise a \emph{NumberFormatException}, like so:

\begin{lstlisting}
void func_2147483648() {
}
\end{lstlisting}

\noindent Since the \emph{Namer} class is used for all front-ends, the issue applied to all languages supported by \V/. It has since been fixed.

\subsection*{Issue 1346: Nonexistent variables in LLVM binops crash \V/}

At the time of testing, \V/ supports a preliminary syntax for embedding method contracts in LLVM assembly language files, with the \emph{VC.contract} syntax.
Referring to nonexistent variables (here `a') in a \emph{requires} clause is correctly handled, except in the case where the variable is used in a LLVM binary operator.
In this scenario, the usage triggers a \emph{NoSuchElementException} and a crash: 

\begin{lstlisting}
define void @example()
!VC.contract !{ !" requires and(%a, %a); " }
{
    ret void
}
\end{lstlisting}

\noindent However, this method contract syntax is slated to be deprecated in the future, in favor of a new contract format that has not been publicly documented yet.
\end{document}